\documentclass[preprint,12pt]{elsarticle}

\usepackage{amssymb}
\usepackage{amsmath,amssymb,amsfonts}
\usepackage[colorlinks=true, linkcolor=blue]{hyperref}
\usepackage{cleveref}
\usepackage{graphicx}
\usepackage{textcomp}
\usepackage{xcolor}
\usepackage{float}
\usepackage{stfloats} 
\usepackage{enumitem}
\usepackage{algorithm}
\usepackage{lineno}
\usepackage{threeparttable}
\usepackage{balance}
\usepackage{algpseudocode}

\begin{document}

\begin{frontmatter}



	\title{Radio Resource Allocation for Beam
		Hopping Scheduling in LEO Satellite
		Communications: A Spatio-Temporal Perspective}


	\author{Hao Yuan$^{a}$, Lanyining Li$^{a}$, Jianghua Long$^{a}$ and Xing Zhang$^{a,*}$}

	\affiliation{
		organization={School of Information and Communication Engineering},
		addressline={Beijing University of Posts and Telecommunications},
		city={Beijing},
		postcode={100876},
		country={China}
	}

\renewcommand{\thefootnote}{\fnsymbol{footnote}}

\footnotetext[1]{Corresponding author.}

\renewcommand{\thefootnote}{\arabic{footnote}}

	\begin{abstract}

        Low Earth Orbit (LEO) satellite networks face critical challenges in radio resource allocation due to dynamic traffic demands and stringent interference constraints. Beam-hopping (BH) technology offers a promising solution by enabling dynamic beam resource allocation across spatial and temporal domains. In this paper, we propose a Tabu Search-based spatio-temporal BH resource allocation strategy for LEO satellite communication systems. Specifically, the BH scheduling problem is formulated to maximize user demand satisfaction under interference constraints. To solve this problem efficiently, the proposed Tabu Search framework integrates adaptive tabu tenure control, greedy-based initialization with interference-aware beam selection, and Simulated Annealing acceptance criteria. Extensive simulation results demonstrate that the proposed method consistently improves system throughput by 17.2\% and user satisfaction by 11.7\% compared with greedy-based BH strategies. These results indicate that the proposed approach provides a scalable and robust solution for dynamic resource allocation in interference-limited LEO satellite networks.
	\end{abstract}

		\noindent\textit{Note:} This paper has been published in \textit{Computer Networks} (Elsevier), 2026.
		The final published version is available at \url{https://doi.org/10.1016/j.comnet.2026.112404}.

	\begin{keyword}


		Beam hopping, Tabu Search, Resource allocation, LEO satellite communications
	\end{keyword}

\end{frontmatter}



\section{Introduction}

With the rapid development of low Earth orbit (LEO) satellite constellations, LEO satellite communications have emerged as a critical infrastructure for enabling global broadband access \cite{Kodheli2021, Xiao2024}. Compared with geostationary orbit (GEO) systems, LEO satellites offer advantages including lower communication latency, reduced link loss, and higher spectrum reuse efficiency. Thus, these characteristics make LEO satellites particularly well-suited for remote access devices, emergency communications, and Internet of Things (IoT) backhaul applications that require low latency and dynamic coverage \cite{Wang2023, Hoyhtya2022}. However, the constrained transmission resources and the dynamic distribution of ground users present significant challenges for resource scheduling, which remains a critical bottleneck to further improving system performance \cite{Yue2023}.

Beam hopping (BH) is a dynamic resource allocation technique \cite{Wang2023electronics} that performs spatio-temporal scheduling by dynamically switching beams across cells and time slots. In contrast to static beam coverage, BH dynamically adjusts both the coverage area and time allocation based on ground user demand, thereby improving spectrum efficiency and beam resource utilization. This is especially effective in scenarios with non-uniform user distributions and highly dynamic spatio-temporal traffic demand \cite{Zhao2025}.

In practical systems, beam-level spatio-temporal hopping faces several key challenges. These include high-dimensional discrete combinatorial optimization, limited time slot resources, and interference coordination. The BH scheduling problem is fundamentally a discrete combinatorial optimization task. Its complexity arises from jointly optimizing beam activation sequences and managing interference \cite{Shao2024}. In addition, under the constraints of system transmit power, bandwidth, and quality of service (QoS) requirements, it is also necessary to balance the objectives of maximizing throughput and ensuring coverage fairness \cite{Meng2024}. Moreover, due to the continuous changes in satellite on-orbit movement and ground service demand, the BH scheduling system is required to exhibit high real-time performance and robustness. This imposes strict requirements on the algorithm's response speed and generalization ability \cite{Li2023}.

\subsection{Related Work}

Early beam hopping studies predominantly targeted GEO satellites \cite{alegre2011heuristic}. However, LEO constellations present new challenges due to limited onboard resources and satellite mobility \cite{tang2021optimization}. In recent years, numerous schemes have been proposed to perform dynamic beam hopping in LEO satellite networks. This section reviews representative BH scheduling strategies, including greedy-based, evolutionary optimization, convex optimization, and intelligent learning-based approaches.
\subsubsection{Greedy-Based Scheduling}
Many works start with greedy-based algorithms for spatio-temporal scheduling \cite{zhang2021dynamic}. For instance, Liu et al. demonstrated an iterative demand-based beam scheduling strategy \cite{tang2021resource}. Other researchers explored traffic-driven heuristics such as clustering and greedy allocation \cite{yang2023cluster}. Greedy algorithms are widely adopted due to their low complexity and fast decision-making capability. In \cite{chen2022next}, Chen \textit{et al.} proposed a load-balancing BH strategy that prioritizes beam slots with higher traffic demands while avoiding co-channel interference via beam separation constraints. The approach achieves significant throughput gains under dynamic traffic scenarios. Similarly, Lin \textit{et al.} designed a beam-slot pairing mechanism based on instantaneous traffic weights, demonstrating notable latency reduction in hotspot service areas in \cite{lin2022multi}. Wang \textit{et al.} proposed a dual-mode greedy scheduling framework that integrates a demand-weighted greedy initializer with an offline SCA/DP-based power refinement phase in \cite{Wang2023electronics}. Zhang \textit{et al.} introduced a capacity-aware greedy scheduler under interference constraints in \cite{zhang2023capacity}. However, such methods are prone to local optima and lack fairness across low-demand regions. Greedy‐based BH scheduling algorithms offer a compelling low‐complexity alternative to exhaustive optimization in multi‐beam LEO satellite systems. The greedy‐based beam hopping (G-BH) method selects the top $K$ beam positions with the highest traffic demand per slot and adopts simple average power allocation, enabling fast and scalable scheduling~\cite{Zheng2023_GBH}. In interference-aware environments, ~\cite{han2023beam} adopted the Viterbi algorithm to design regular BH matrices and a greedy algorithm for irregular BH matrices. Simulation results show that the Viterbi‐based regular pattern achieves approximately 2.8 dB transmit power reduction compared with existing baseline schemes, while the irregular pattern further improves performance by about 2 dB relative to the regular pattern. 
\textbf{Despite their practical advantages for real-time and online adaptations, greedy heuristics may be myopic and can struggle to balance throughput, delay, energy efficiency, and fairness across heterogeneous traffic regions. These limitations motivate the investigation of more global search-based optimization methods, among which evolutionary algorithms have emerged as a promising class of solutions for BH scheduling problems.}

\subsubsection{Evolutionary Optimization}
BH scheduling in next-generation high-throughput and LEO satellite systems has increasingly been formulated as a large-scale combinatorial optimization problem, where evolutionary computation offers effective, constraint-aware solutions. Genetic algorithms (GA) have been applied to explore large solution spaces with strong global search capabilities. Early work by Zhang \textit{et al.} \cite{zhang2021dynamic} applied genetic algorithms to dynamic time‑slot allocation in beam hopping satellite systems under time-varying rain attenuation. Simulations demonstrate that, compared to conventional schemes assuming constant attenuation, the GA‑based approach dynamically adapts time‑slot allocation, resulting in significant improvements in overall system performance.
Deng \textit{et al.} applied a GA-based approach to address both throughput-driven and satisfaction-driven BH scheduling schemes, incorporating interference constraints derived from satellite-terrestrial spectrum sharing. Their results demonstrated that GA could efficiently adapt to non-uniform traffic distributions and dynamic fading environments in \cite{deng2024satellites}. These studies underline the flexibility of GA in modeling mixed-integer decision spaces and optimizing multiple conflicting objectives under stochastic conditions. Subsequent interference-aware GA variants for LEO BH systems achieved significant reductions in co-channel interference while maintaining service fairness in \cite{panpan42beam}. Particle swarm optimization (PSO) has been employed for beam-to-slot assignment and for the joint allocation of time, frequency, and power, offering rapid convergence and flexible constraint handling \cite{zhang2025particle}. Emerging research directions include hybrid evolutionary learning approaches, scaling to multi-satellite and multi-resource coupling, and incorporating operator constraints such as GEO-LEO coexistence across full BH cycles \cite{guo2022efficient}.

\subsubsection{Convex Optimization-Based Scheduling}
In recent years, BH scheduling in high-throughput multi-beam satellite systems has often been formulated as mixed-integer and non-convex optimization problems. A promising approach is to decouple the combinatorial scheduling stage from the continuous power or bandwidth allocation stage by leveraging convex-optimization techniques, such as successive convex approximation (SCA), conic programming, and dual decomposition. Convex programming has formulated BH scheduling as a multi-objective resource allocation problem. Zhou \textit{et al.} \cite{zhou2024multi} developed a selective precoding approach with convex relaxation, enabling global optimality under power and interference constraints. Jia \textit{et al.} \cite{jia2024dynamic} proposed a beam-hopping and resource allocation algorithm tailored for NGSO satellite systems facing non-uniform traffic demand. Leveraging Lyapunov optimization, the authors transform the original stochastic formulation into a deterministic control approach that jointly determines beam hopping patterns, bandwidth allocation, and power control in an online manner. However, the effectiveness of these models depends heavily on accurate channel and traffic estimation, which may be impractical in real-time satellite environments. Chen \textit{et al.} introduced a two-stage framework that models beam scheduling probabilistically and handles power allocation via convex optimization, achieving high throughput and reduced energy consumption under heterogeneous service demands \cite{chen2024joint}. Wang \textit{et al.} further showed that joint BH and NOMA scheduling is NP-hard, and leveraged mixed-integer conic programming (MIPC) to derive tight upper/lower bounds and efficient heuristics that approximate the original non-convex problem \cite{wang2022joint, wang2021joint}. In integrated satellite and terrestrial or multi-satellite scenarios, newer works embed convex subproblems for inter-beam power control within global scheduling routines, maintaining QoS, fairness, and energy objectives with polynomial complexity \cite{zhang2025framework}. A study on cooperative satellite systems introduces a coordinated load-balancing and BH scheduling framework that follows the paradigm of convex continuous resource optimization combined with combinatorial scheduling \cite{wang2024resource}. Other investigations applied the ADMM algorithm and bisection methods to alternating-optimize BH patterns or grant-free access in LEO networks, further demonstrating the versatility of convex optimizers for pattern design \cite{jeon2025beam}. Overall, convex optimization tools, including SCA, conic programming, and dual methods, serve as effective decoupling mechanisms that transform complex rate–interference coupling and QoS-constrained BH scheduling into tractable iterative subproblems, while providing reproducible baselines for future satellite communication systems~\cite{shen2018fractional}.

\subsubsection{Learning-Based Approaches}
With the continuous development of artificial intelligence, reinforcement learning (RL) has been widely applied in dynamic resource management and interference mitigation \cite{jiang2020reinforcement}. In particular, multi-agent reinforcement learning (MARL), which involves collaboration and competition among multiple agents, has the ability to find efficient solutions in complex environments \cite{holder2025multi}. MARL has significant advantages in addressing resource allocation and interference mitigation in LEO satellite systems, as it allows information sharing between satellites and the real-time adjustment of decision-making strategies based on environmental changes. To reduce inference latency, Kim \textit{et al.} developed a lightweight DQN-based scheduler with compressed state representations, suitable for onboard deployment \cite{kim2025dqn}.

Benefiting from the outstanding performance of deep reinforcement learning (DRL) in high-dimensional decision-making tasks, researchers have increasingly explored its application in satellite communication systems, particularly through single-agent frameworks. Algorithms such as Deep Q-Network (DQN), Proximal Policy Optimization (PPO), and Soft Actor-Critic (SAC) have been employed to tackle key problems, including beam hopping, power control, and dynamic routing, leading to significant improvements in both link capacity and spectrum efficiency \cite{zhang2025deep, xie2025multi, giraldo2024deep}.

For instance, the DRL-based Beam Hopping Scheduling (DRL-BHS) algorithm is capable of adaptively selecting beam illumination patterns based on dynamic traffic heatmaps, achieving approximately 25\% throughput gains over static beam hopping schemes \cite{zhang2025deep}. Moreover, joint optimization of beam hopping and power allocation using PPO has demonstrated superior interference suppression compared to traditional greedy algorithms in non-geostationary satellite orbit (NGSO) constellations \cite{xie2025multi}. Additionally, the integration of Graph Neural Networks (GNNs) with DRL has shown initial promise in modeling and managing complex network topologies, enabling more effective and context-aware resource allocation strategies \cite{chen2025deep}. However, DRL-based methods require extensive training data and computational resources, and their generalization remains a challenge.

\subsection{Motivation and Contributions}

Nevertheless, these methods still face several practical limitations. Traditional heuristic algorithms typically rely on fixed neighborhood search strategies and parameter settings, rendering them prone to local optima and less effective in multi-objective, highly constrained, or multimodal optimization scenarios \cite{Xu2025}. Although intelligent algorithms have end-to-end learning capabilities, they often require large amounts of interaction data, incur high computational costs, and have slow convergence. Furthermore, their generalization ability may be limited when faced with environmental changes, including orbital dynamics or variations in traffic demand models \cite{Zamacola2025}. Finally, the black-box nature of such models poses challenges for interpretability and stability in real-world deployments \cite{Cui2025}.

To address these challenges, this paper introduces several novel contributions. Firstly, a spatio-temporal BH resource allocation approach is designed by introducing a problem-size-adaptive tabu tenure control mechanism. Combined with a probabilistic acceptance of worse solutions, this strategy achieves optimal BH scheduling that balances user coverage and dynamic traffic demand. Secondly, a two-stage solution generation and evaluation framework is constructed: the first stage generates an initial solution by prioritizing cells with higher residual traffic demand while respecting intra-satellite interference distance constraints; the second stage designs a traffic demand satisfaction evaluation function and performs neighborhood exploration and solution selection tailored to different objectives. This ensures a balance between local optimization efficiency and global search capability under satellite power and bandwidth resource constraints.

The structure of this paper is as follows. Section II shows the system model, including the multi-beam satellite system, user traffic model, and formulates the BH scheduling problem as a constrained optimization task. Section III details the proposed  Tabu Search-based BH scheduling procedure. Section IV presents simulation results and comparative analysis to validate the performance of the proposed method. Finally, the paper ends with a summary in Section V.

\section{System Model and Problem Formulation}

\subsection{BH LEO Satellite Communication Systems}

This subsection describes the basic principle and parameter modeling of BH in LEO satellite communication systems. Fig. 1 illustrates the forward‑link BH LEO system scenario, where service beams dynamically hop among predefined cells to improve ground coverage and reduce idle resource waste.

\begin{figure*}[htbp]
	\centering
	\includegraphics[width=\textwidth,keepaspectratio]{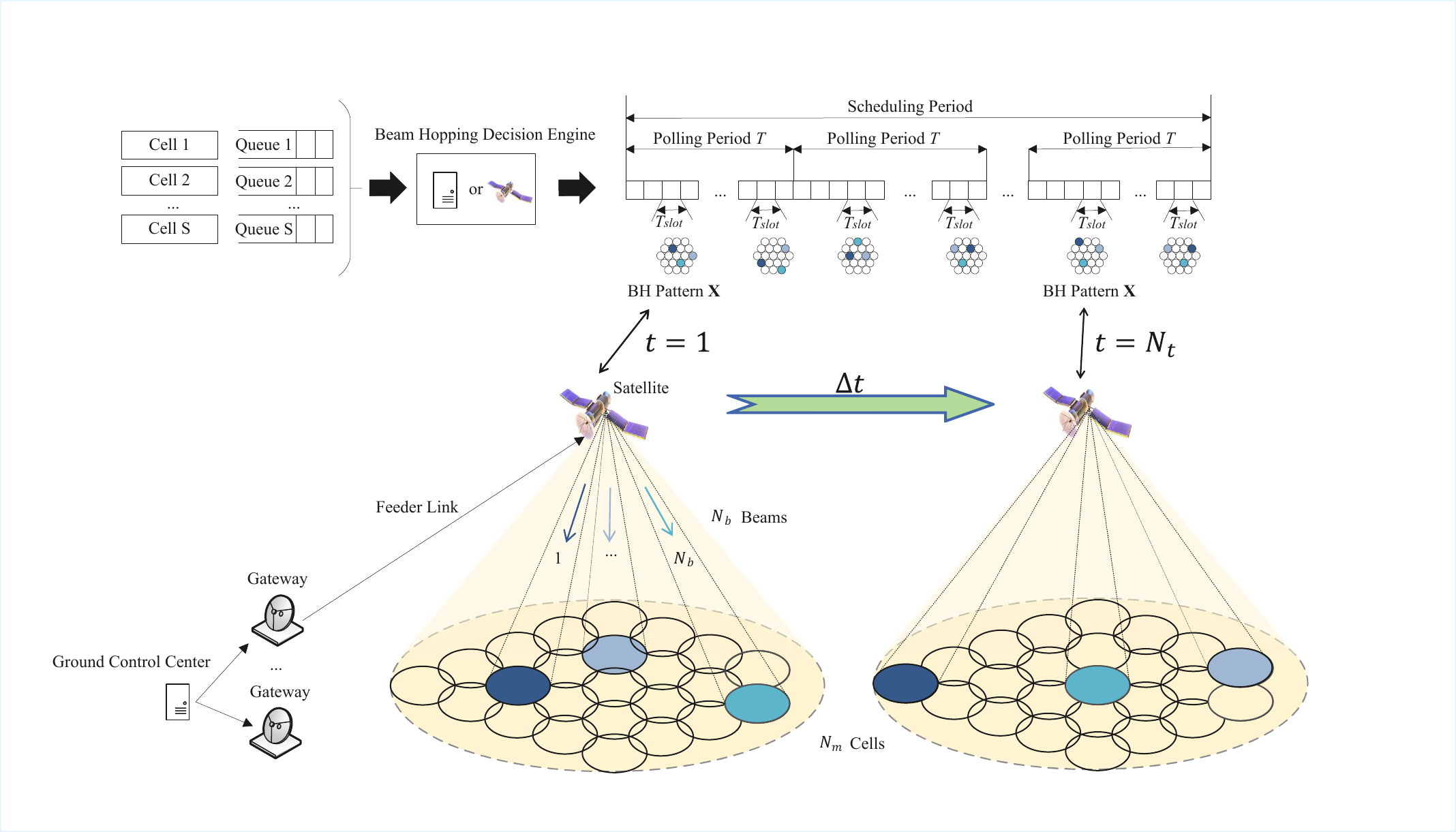}
	\caption{Forward-Link BH LEO Satellite Communication Systems.}
	\label{fig:system-crop}
\end{figure*}

The satellite coverage area is discretized into a set of cells defined as
\begin{equation}
	\mathcal{M} = \{ m_1, m_2, \ldots, m_{N_m} \},
\end{equation}
where \(N_m\) denotes the total number of cells. The set of users covered by cell $m_i$ is denoted as
\begin{equation}
	\mathcal{U}_{m_i} = \{u_1, u_2, \ldots, u_{K_i}\},
\end{equation}
where $m_i$ represents the $i$-th cell with $i \in \{1, 2, \ldots, N_m\}$, $K_i$ represents the number of users located within the coverage area of cell $m_i$.
The scheduling period is divided into time slots
\begin{equation}
	t \in \{1, 2, \ldots, N_t\},
\end{equation}
during which the satellite can activate \(N_b\) beams, with \(N_b \leq N_m\) representing the maximum number of simultaneously active beams. We assume that all active beams share the total onboard bandwidth \(B\) and the total transmit power \(P_s\). Each beam is allocated $P_s / N_b$ of the total power.

The downlink channel model includes free space path loss and rain attenuation:

\begin{equation}
L_{total} = L_{FS} + L_{rain}
\end{equation}

where $L_{\mathrm{FS}}$ denotes the free-space path loss calculated at the carrier frequency $f = 3.62\,\mathrm{GHz}$ and slant range $d$, which varies from $550$ to $800\,\mathrm{km}$ depending on the elevation angle. For nadir pointing ($d = 550\,\mathrm{km}$), the free-space path loss is $L_{\mathrm{FS}} = 162.5\,\mathrm{dB}$. The rain attenuation, denoted by $L_{\mathrm{rain}}$, is modeled according to ITU-R P.618, with a fade margin of $2$--$5\,\mathrm{dB}$ for Ka-band operation in temperate climates under $99.9\%$ availability.

The satellite antenna is modeled as a phased array with a peak gain of $G_{\mathrm{sat}} = 30\,\mathrm{dBi}$ and a $3\,\mathrm{dB}$ beamwidth of $\theta_{3\mathrm{dB}} = 4.5^\circ$. User terminals are assumed to employ omnidirectional antennas with $G_{\mathrm{user}} = 0\,\mathrm{dBi}$.

To model the time-domain activation pattern of BH in a scheduling period, we define a binary scheduling matrix that captures the spatio-temporal resource allocation across cells and time slots.

Specifically, the BH schedule is described by a binary matrix of size $N_m \times N_t$, which can be given by
\begin{equation}
	\boldsymbol{X} = \left\{ x_{m_i,t} \;\middle|\; m_i \in \mathcal{M},\;
	t \in \{1,2,\ldots,N_t\} \right\},
	\label{eq:setX}
\end{equation}
where  $x_{m_i,t} = 1$ indicates that cell $m_i$ is activated in time slot $t$; $x_{m_i,t} = 0$ indicates that cell $m_i$ is not activated in time slot $t$.

Additionally, from matrix $\boldsymbol{X}$, allocation statistics can be extracted for each cell over the scheduling period. Specifically, the total number of time slots allocated to the cell $m_i$ is defined as
\begin{equation}
	T_{m_i} = \sum_{t=1}^{N_t} x_{m_i,t},
	\label{eq:Tmi}
\end{equation}
which serves as a key metric to evaluate whether the resource allocation adequately matches the traffic demand of the cell. A higher value of $T_{m_i}$ indicates more frequent activation, which can be tailored to satisfy non-uniform traffic distributions of the service area.

This matrix representation provides a flexible framework for formulating the BH scheduling problem as a constrained optimization task and enables the joint design of activation patterns, interference management, and QoS provisioning in LEO satellite communication systems.

\subsection{User Traffic Model}

The user behavior in this study is modeled using the File Transfer Protocol (FTP) traffic model, which accurately captures the bursty and intermittent characteristics of non-real-time services such as file downloads. Specifically, the file download request events of an individual user $u$ follow a Poisson process with rate parameter $\lambda_u$. This implies that the number of requests within a unit time interval is an independent random variable, with the average request rate $\lambda_u$ fluctuating dynamically within the interval $[\lambda_{\text{min}}, \lambda_{\text{max}}]$ due to factors such as time-of-day or regional user density variations.

The traffic demand of a single request is modeled as an exponentially distributed random variable denoted by $R_u$ (unit: bits). It represents the traffic demand that a user $u$ needs to transmit per request. The spatio-temporal distribution of user demand is a core input for BH scheduling design, as ground users typically exhibit significant spatial heterogeneity and stochasticity in their communication needs.

Given that users within the same cell share the beam resources, the cell $m_i$ is regarded as the fundamental unit of resource allocation. The total traffic demand of the cell can be expressed by
\begin{equation}
	R_i = \sum_{u \in \mathcal{U}_{m_i}} R_u,
	\label{eq:Ri}
\end{equation}
where $\mathcal{U}_{m_i} = \{u_1, u_2, \ldots, u_{K_i}\}$ denotes the set of users covered by the cell, and $K_i$ is the total number of users in the cell.
The scheduling period $T$ is divided into $N_t$ time slots, each with a duration of $\Delta t = 0.5\,\mathrm{ms}$. For a scheduling period of $T = 40\,\mathrm{ms}$, the total number of time slots is therefore
\[
N_t = \frac{T}{\Delta t} = \frac{40}{0.5} = 80.
\]

If the cell $m_i$ is activated for a total duration of
\begin{equation}
	T_i = \sum_{t=1}^{N_t} x_{m_i,t} \cdot \Delta t
	\label{eq:Ti}
\end{equation}
over the scheduling period, where $\Delta t$ is the duration of a single time slot, its average transmission rate can be calculated by \( R_i / T_i \) (bps).

The interference threshold \(D_{\mathrm{thr}}\) is defined as the minimum center-to-center distance between simultaneously active beams required to limit co-channel interference. In this work, we set \(D_{\mathrm{thr}} = 2.0R_b = 100~\mathrm{km}\), where \(R_b = 50~\mathrm{km}\) denotes the beam radius. Under this setting, the interference introduced by the nearest simultaneously active beam results in less than \(1~\mathrm{dB}\) SINR degradation, which is considered acceptable for the targeted service scenarios.

\subsection{Performance Metrics}
To comprehensively evaluate the proposed scheduler from a user-centric perspective, we define the following Key Performance Indicators (KPIs):

\subsubsection{Service Satisfaction}

For user $i$, the service satisfaction is defined as:
\begin{equation}
SS_i = \min\left(1, \frac{\sum_{t=1}^{N_t} \Delta t \cdot x_{m_i,t} \cdot C_i}{R_i}\right)
\label{eq:satisfaction}
\end{equation}

The average service satisfaction and service satisfaction rate are:
\begin{align}
SS_{\text{avg}} &= \frac{1}{N} \sum_{i=1}^{N} SS_i \\
SSR_\theta &= \frac{|\{i : SS_i \geq \theta\}|}{N} \times 100\%
\end{align}

where $\theta = 0.9$ defines the threshold for ``satisfied users.'' The $SSR_{0.9}$ metric
directly relates to user QoE, as users with $SS \geq 0.9$ experience good or excellent
service quality for typical applications such as video streaming and web browsing.

\subsubsection{User-centric Throughput Distribution}
Beyond average throughput, we analyze the cumulative distribution function (CDF) of user throughput to evaluate fairness. Specifically, we report:
\begin{itemize}
	\item \textbf{50th Percentile (p50) Throughput:} Represents the median data rate, indicating the typical user experience.
	\item \textbf{95th Percentile (p95) Throughput:} Represents the peak experience for users with the best channel conditions.
\end{itemize}



\subsection{Problem Formulation}

To optimize the BH schedule, it is assumed that users within each cell equally share the allocated power and frequency resources and that their spatial locations remain fixed throughout the scheduling period. Thus, the achievable transmission rate for the cell $m_i$ remains constant over time slots and is denoted by $C_i$.


To maximize traffic demand satisfaction, we aim to minimize the gap between the total supplied capacity and the user demand over the entire scheduling period. The objective function is formulated as follows:
\begin{equation}
	\min \sum_{i=1}^{N_m} \left| \left( \sum_{t=1}^{N_t} \Delta t \cdot x_{m_i,t} C_i \right) - R_i \right|^2
	\label{eq:objective}
\end{equation}

where \( \Delta t \) denotes the duration of each time slot, \( x_{m_i,t} \) is a binary indicator representing whether cell \( m_i \) is active in time slot \( t \), \( C_i \) denotes the transmission rate allocated to cell \( m_i \), and \( R_i \) represents the traffic demand of cell \( m_i \).

The optimization is subject to the following constraints. First, in each time slot, the total number of active beams is limited to $N_b$, which can be calculated as
\begin{equation}
	0 \leq \sum_{i=1}^{N_m} x_{m_i,t} \leq N_b,\quad \forall t \in \{1,\ldots,N_t\}.
	\label{eq:constraint1}
\end{equation}
Additionally, the elements of the BH schedule matrix are binary variables, which can be formulated as
\begin{equation}
	x_{m_i,t} \in \{0,1\}, \quad \forall m_i,\, t.
	\label{eq:constraint2}
\end{equation}

\section{Tabu Search-Based BH Scheduling Scheme}

The proposed scheme comprises the following steps (see Algorithm 1). To construct an initial feasible solution for Tabu Search, a greedy heuristic algorithm that accounts for interference distance constraints is adopted to guide the scheduling process. The objective is to maximize the total serviced traffic demand within a time slot while avoiding interference among simultaneously active beams.

\textbf{Step 1: Initialization and Selection Process}

\begin{itemize}
	\item \textbf{Priority Metric Based on Residual Demand:} Cells are first ranked according to their residual traffic demand $R_i^*$. The priority score for cell $m_i$ is defined as
	      \begin{equation}
		      \mathit{Pri}(m_i) = \frac{R_i^*}{\sum_{m_i \in \mathcal{M}} R_i^*}
		      \label{eq:priority}
	      \end{equation}

	\item \textbf{Interference Distance Constraint:} The two cells $m_i$ and $m_k$ are subject to mutual interference if their geographic distance satisfies $d(m_i, m_k) < D_{\text{thr}}$, where $D_{\text{thr}}$ is a predefined interference threshold. Such cells cannot be activated simultaneously.

	\item \textbf{Greedy-Based Initial Solution Generation:} Starting from the highest-priority cell, the algorithm iteratively selects cells that do not violate the interference distance constraint. This process continues until the maximum number of active beams $N_b$ is reached. The resulting initial solution is
	      \begin{equation}
		      x_0 = \{m_1, m_2, \ldots, m_{N_b}\}.
		      \label{eq:init_sol}
	      \end{equation}

\end{itemize}

To maximize the total priority of selected cells while satisfying spatial separation and beam activation constraints, the initial beam activation set $x_0$ is determined by solving the following optimization problem, which can be expressed by

\begin{align}
	\text{max}\quad
	 & \sum_{m_i \in x_0} \mathit{Pri}(m_i) \label{eq:init_obj}                                                                                            \\[2pt]
	\text{s.t.}\quad
	 & d(m_i, m_k) \geq D_{\text{thr}},                         &  & \forall\, m_i, m_k \in x_0,\; m_i \neq m_k \tag{\ref{eq:init_obj}a}\label{eq:init_c1} \\[2pt]
	 & |x_0| \leq N_b,                                          &  & \tag{\ref{eq:init_obj}b}\label{eq:init_c2}
\end{align}
where $\mathit{Pri}(m_i)$ denotes the priority score of cell $m_i$, which is typically derived from residual traffic demand. Constraint~\eqref{eq:init_c1} ensures that any two selected beams are separated by at least the interference threshold $D_{\text{thr}}$, while \eqref{eq:init_c2} limits the total number of activated beams to $N_b$.

This initialization ensures that the selected set of cells maximizes the weighted sum of residual demand while strictly satisfying interference avoidance constraints.

\textbf{Step 2: Tabu List Design}

The tabu list is designed to record historical cell activation patterns over time, thereby preventing redundant searches and cycling during the optimization process. Its structure is defined as a two-dimensional matrix of size $N_b \times L_{\text{tabu}}$, where each row corresponds to a beam position and each column stores the indices of cells that were recently selected for service.

Fig. 2 illustrates the conceptual design of the tabu list. Each cell in the matrix represents a cell index associated with a specific activation history. During each iteration of the optimization, newly selected candidate solutions are incorporated into the tabu list to enforce search diversification.

\begin{figure}[htbp]
	\centering
	\includegraphics[width=\linewidth,keepaspectratio]{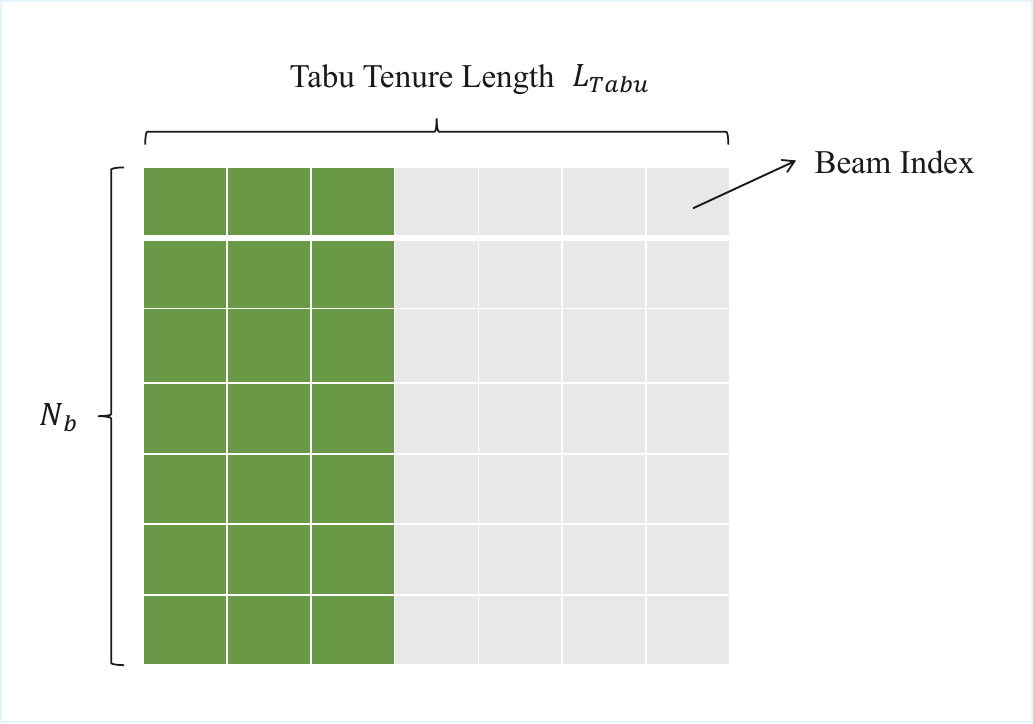}
	\caption{Tabu List Design: Dynamic Tabu Tenure Parameter Configuration.}
	\label{fig:tabu_list_design}
\end{figure}

The tabu tenure length $L_{\text{tabu}}$ is dynamically configured based on the number of satellite beams $N_b$ and the total number of cells $N_m$ to balance exploration and exploitation. The adaptive formulation is given by:
\begin{equation}
	L_{\text{tabu}} = \lfloor \sqrt{N_b} \cdot \sqrt{N_m} \rfloor
	\label{eq:ltabu}
\end{equation}
This adaptive formulation scales with the problem size, ensuring sufficient diversity in the search while avoiding excessive restriction. For the typical scenario with $N_b =10$ and $N_m=50$, this yields $L_{\text{tabu}} \approx 22$, providing an effective balance between memory capacity and search flexibility.

\textbf{Step 3: Neighborhood Move Procedure}

In this step, the algorithm generates a specified number of neighborhood solutions. Specifically, the random beam count + random beam index strategy was used to construct neighborhood moves.

First, a random number $K \in [1, N_b]$ is generated to determine how many cells will be replaced. Then, positions within the current solution $x_{\text{now}}$ are randomly selected from
\begin{equation}
	\{p_1, p_2, \ldots, p_K\}.
	\label{eq:priority_set}
\end{equation}
For each selected position, cells from the set of non-active candidates $M \setminus x_{\text{now}}$ are randomly chosen to replace them. The resulting neighborhood solution $x_{\text{candidate}}$ is defined as
\begin{equation}
	x_{\text{candidate}} = (x_{\text{now}} \setminus \{x_{\text{now}}[p_k]\}) \cup \{m_{p_k}\},
	\label{eq:xcandidate}
\end{equation}
where $x_{\text{now}}[p_k]$ denotes the cell at position $p_k$ in the current solution.

Fig. 3 illustrates the neighborhood move procedure used in the Tabu Search algorithm.
\begin{figure}[htbp]
	\centering
	\includegraphics[width=\linewidth,keepaspectratio]{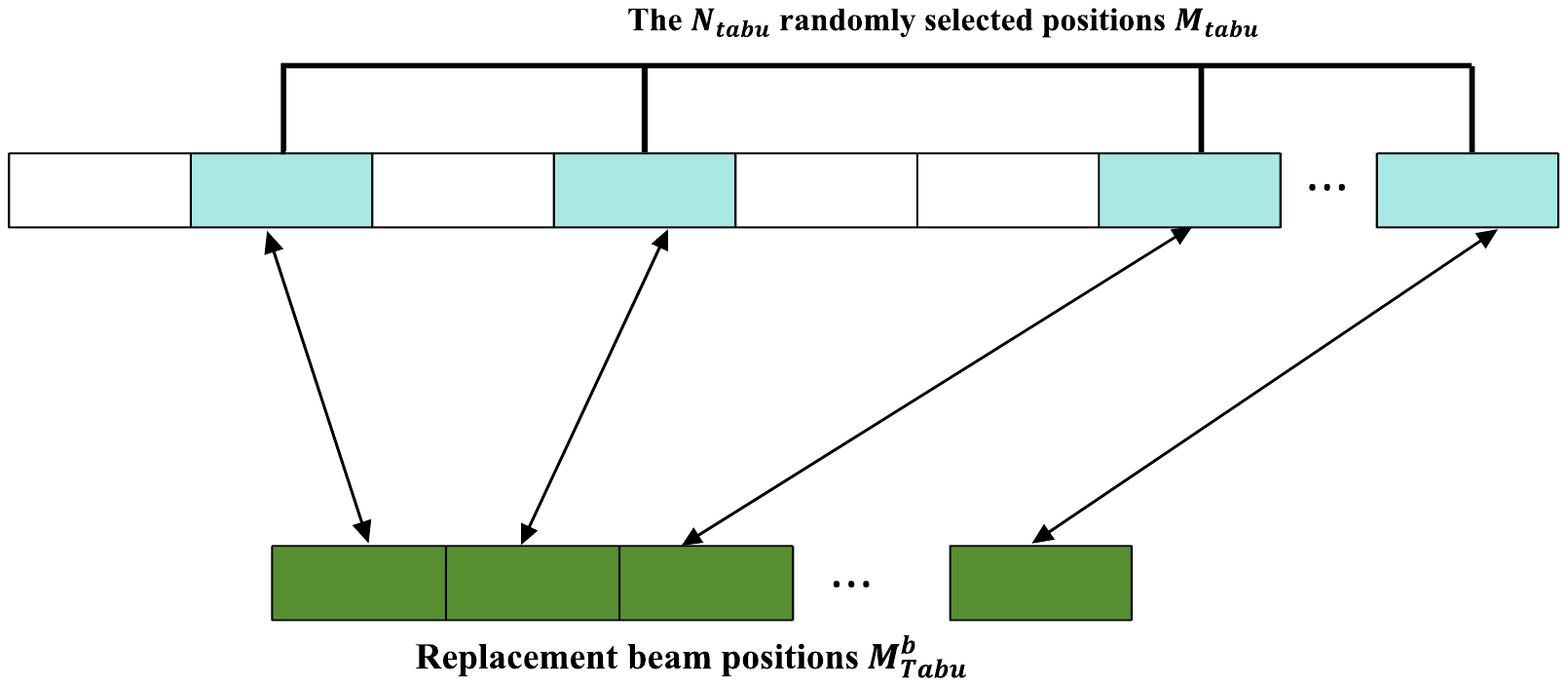}
	\caption{Neighborhood Move Procedure.}
	\label{fig:neighborhood_move}
\end{figure}

To improve convergence efficiency, the algorithm incorporates problem-specific heuristics to guide neighborhood generation.

\begin{itemize}
	\item \textbf{Priority-Based Neighborhood Generation:} Cells in the current solution are ranked by residual demand $R_i^*$. Lower-priority cells are preferred for replacement to ensure that the overall priority of the generated neighborhood solution does not decrease.

	\item \textbf{Interference-Avoidance Neighborhood Generation:} Replacement candidates are selected from non-active cells that satisfy constraint \eqref{eq:init_c1}.

\end{itemize}


\textbf{Step 4: Evaluation and Acceptance Strategy}

In this step, the algorithm evaluates the best candidate solution, denoted as $x_{candidate}$, found in the current neighborhood. To balance exploration (avoiding local optima) and exploitation (refining the best solution), we integrate a Tabu restriction with a Simulated Annealing (SA) probabilistic acceptance criterion.
The evaluation function is defined as

\begin{equation}
	f(x)= \sum_{i=1}^{N_m} \left| \left( \sum_{t=1}^{N_t} \Delta t \cdot x_{m_i,t} C_i \right) - R_i \right|^2
	\label{eq:eval_func}
\end{equation}

The detailed acceptance logic is as follows:

\begin{itemize}
	\item \textbf{Tabu Check:} First, we verify if the move generating $x_{candidate}$ exists in the Tabu List.
	      \begin{itemize}
		      \item If the move is \textbf{Tabu} and does not satisfy the aspiration criterion (i.e., $E(x_{candidate}) \ge E(x_{best})$), the candidate is rejected.
		      \item If the move is \textbf{Non-Tabu} (or satisfies the aspiration criterion), we proceed to the energy evaluation.
	      \end{itemize}

	\item \textbf{Energy Evaluation and SA Criterion:} We calculate the energy difference $\Delta E = E(x_{candidate}) - E(x_{now})$.
	      \begin{itemize}
		      \item \textbf{Improvement ($\Delta E < 0$):} If the candidate yields a lower objective value, it is accepted immediately, and we update $x_{now} \leftarrow x_{candidate}$.
		      \item \textbf{Deterioration ($\Delta E \ge 0$):} If the candidate is worse than the current solution, we employ the Metropolis criterion from Simulated Annealing. The acceptance probability $P$ is calculated as:
		            \begin{equation}
			            P = \exp\left( - \frac{E(x_{candidate}) - E(x_{now})}{T_k} \right)
		            \end{equation}
		            A random number $r \in [0, 1]$ is generated. If $r < P$, the worse solution $x_{candidate}$ is accepted to help the search escape local optima; otherwise, it is discarded.
	      \end{itemize}
\end{itemize}

The temperature $T_k$ is updated via an exponential decay schedule $T_{k+1} = \alpha T_k$ to gradually reduce the acceptance probability of worse solutions as the algorithm converges.









\textbf{Step 5: Solution Update}

The step repeats Steps 4.3 to 4.5 until the maximum number of iterations $G_{max}$ is reached. At the end of the iteration process, $x_{\text{best}}$ represents the best cell activation plan. Its cell indices are written into the corresponding row of the BH schedule matrix.

\textbf{Step 6: Repeat}

This step repeats the above process for all time slots $t = 1, 2, \ldots, N_t$ to generate the complete BH schedule matrix $X \in \mathbb{R}^{N_m \times N_t}$. In matrix $X$, $x_{m_i,t} = 1$ indicates that the cell $m_i$ is active in the time slot $t$. This matrix can be directly uploaded to the satellite BH decision engine to implement BH scheduling.
\begin{nolinenumbers}
	\begin{algorithm}[H]
		\caption{Tabu Search-Based BH Scheduling Scheme}
		\textbf{Input:} Number of beams $N_b$, set of cells $\mathcal{M}$, total time slots $N_t$, Tabu Search parameters\\
		\textbf{Output:} BH schedule matrix $X$
		\begin{algorithmic}[1]
			\State Initialize time slot counter $t \gets 0$
			\While{$t \leq N_t$}
			\State Generate initial solution $x_0$
			\State Set $x_{\text{now}} \gets x_0$, $x_{\text{best}} \gets x_0$
			\For{$i = 1$ to $G_{max}$}
			\State Generate $N_{\text{Tabu}}$ neighborhood solutions
			\State Select the best neighbor from the neighborhood as $x_{candidate}$
			\State Calculate energy difference $\Delta E = E(x_{candidate}) - E(x_{now})$

			\If {($x_{candidate}$ is NOT in Tabu List) \textbf{or} ($E(x_{candidate}) < E(x_{best})$)}
			\If {$\Delta E < 0$}
			\State $x_{now} \leftarrow x_{candidate}$ \Comment{Accept Improvement}
			\If {$E(x_{now}) < E(x_{best})$}
			\State $x_{best} \leftarrow x_{now}$
			\EndIf
			\Else
			\State Calculate acceptance probability $P = \exp(-\Delta E / T_k)$
			\State Generate random number $r \sim U(0,1)$
			\If {$r < P$}
			\State $x_{now} \leftarrow x_{candidate}$ \Comment{Accept worse solution by SA}
			\EndIf
			\EndIf
			\State Update Tabu List (add current move, remove expired)
			\EndIf
			\State $T_{k+1} \leftarrow \alpha \cdot T_k$ \Comment{Cooling schedule}
			\EndFor
			\State Assign $x_{\text{best}}$ to the $t$-th column of $X$
			\State $t \gets t + 1$
			\EndWhile
		\end{algorithmic}
	\end{algorithm}
\end{nolinenumbers}

\section{Simulation Results}

The simulation scenario parameters are listed in Table~I.

\begin{table}[htbp]\footnotesize
	\centering
	\begin{threeparttable}
		\setlength{\abovecaptionskip}{0pt}
		\caption{Simulation Parameters}
		\begin{tabular}{p{5cm} p{3.5cm}}
			\hline
			\textbf{Parameter}                 & \textbf{Value}                  \\
			\hline
			\multicolumn{2}{l}{\textit{Satellite Parameters}}                    \\
			Orbital altitude                   & 508 km                          \\
			Orbital inclination                & 53$^\circ$                      \\
			Number of orbital planes           & 60                              \\
            Number of beams           &   10                              \\
			Satellites per orbital plane       & 90                              \\
            Reuse pattern   &   Full frequency reuse                           \\
			\hline
			\multicolumn{2}{l}{\textit{Coverage Parameters}}                     \\
			Coverage longitude range           & [102$^\circ$, 108$^\circ$]      \\
			Coverage latitude range            & [26$^\circ$, 30$^\circ$]        \\
			Beam radius $R_b$                  & 50 km                           \\
            Number of cells $N_m$                  & 50                           \\
			3dB beam angle $\theta_{3dB}$      & 4.5$^\circ$                     \\
			Beam scan range                    & [33$^\circ$, 45$^\circ$]        \\
			Interference threshold $D_{thr}$   & 2.0 $\times R_b$                \\
			\hline
			\multicolumn{2}{l}{\textit{Time Slot Parameters}}                    \\
			Time slot duration $\Delta t$      & 0.5 ms                          \\
			Scheduling period length           & 40 ms                           \\
			Time slots per period $N_t$        & 80                              \\
			Number of scheduling periods       & 25                              \\
			\hline
			\multicolumn{2}{l}{\textit{Frequency Parameters}}                    \\
			Frequency band                     & S-band                          \\
			Downlink frequency $f_{DL}$        & 3.62 GHz                        \\
			Bandwidth $B$                      & 40 MHz                          \\
			Subcarrier spacing $\Delta f$      & 30 kHz                          \\
			\hline
			\multicolumn{2}{l}{\textit{Power Parameters}}                        \\
			Total beam power $P_s$             & 300 W (54.8 dBm)                \\
			Satellite transmit antenna gain    & 30 dBi                          \\
			User terminal receive antenna gain & 0 dBi                           \\
			System noise temperature           & 150 K                           \\
			\hline
			\multicolumn{2}{l}{\textit{Traffic Parameters}}                      \\
			Total number of users              & 800                             \\
			Traffic type                       & FTP                             \\
			File size distribution             & Exponential ($\lambda$=10)   \\
			\hline
			\multicolumn{2}{l}{\textit{Algorithm Parameters}}                    \\
			Maximum iterations $G_{max}$       & 50                              \\
			Neighborhood size $|N(x)|$         & 10                              \\
			Tabu tenure $L_{\text{tabu}}$             & Adaptive (Eq.~\eqref{eq:ltabu}) \\
			Initial temperature $T_0$          & 1000                            \\
			Cooling rate $\alpha$              & 0.95                            \\
			\hline
		\end{tabular}
		\begin{tablenotes}
			\item The interference threshold $D_{thr}$ is set to twice the beam radius to ensure sufficient spatial separation.
			\item The tabu tenure is dynamically calculated based on problem size.
		\end{tablenotes}
	\end{threeparttable}
	\label{tab:simulation-parameters}
\end{table}

The benchmark algorithms used for comparison include the greedy beam-hopping with interference control (GBH-AIC) \cite{Zhang2023BeamHoppingWCNC}, the greedy beam-hopping without interference control (GBH-WIC), and a Genetic Algorithm (GA) baseline. GBH-AIC selects beam spots with the highest residual traffic demand in each time slot while ensuring that the geographic distances between activated beams exceed a predefined interference threshold. In contrast, GBH-WIC activates only the beam spots with the highest residual demand, without considering interference constraints. The GA baseline employs tournament selection, uniform crossover, and swap mutation over a population of 30 individuals for 50 generations, using the same objective function and interference constraints as the proposed method. This provides a representative comparison against evolutionary optimization, which is widely used in satellite resource allocation literature. The performance of these strategies is assessed under a service-deficit scenario, where the available system resources cannot fully accommodate user demands.

\begin{figure*}[!htbp]
	\centering
	\includegraphics[width=0.85\textwidth]{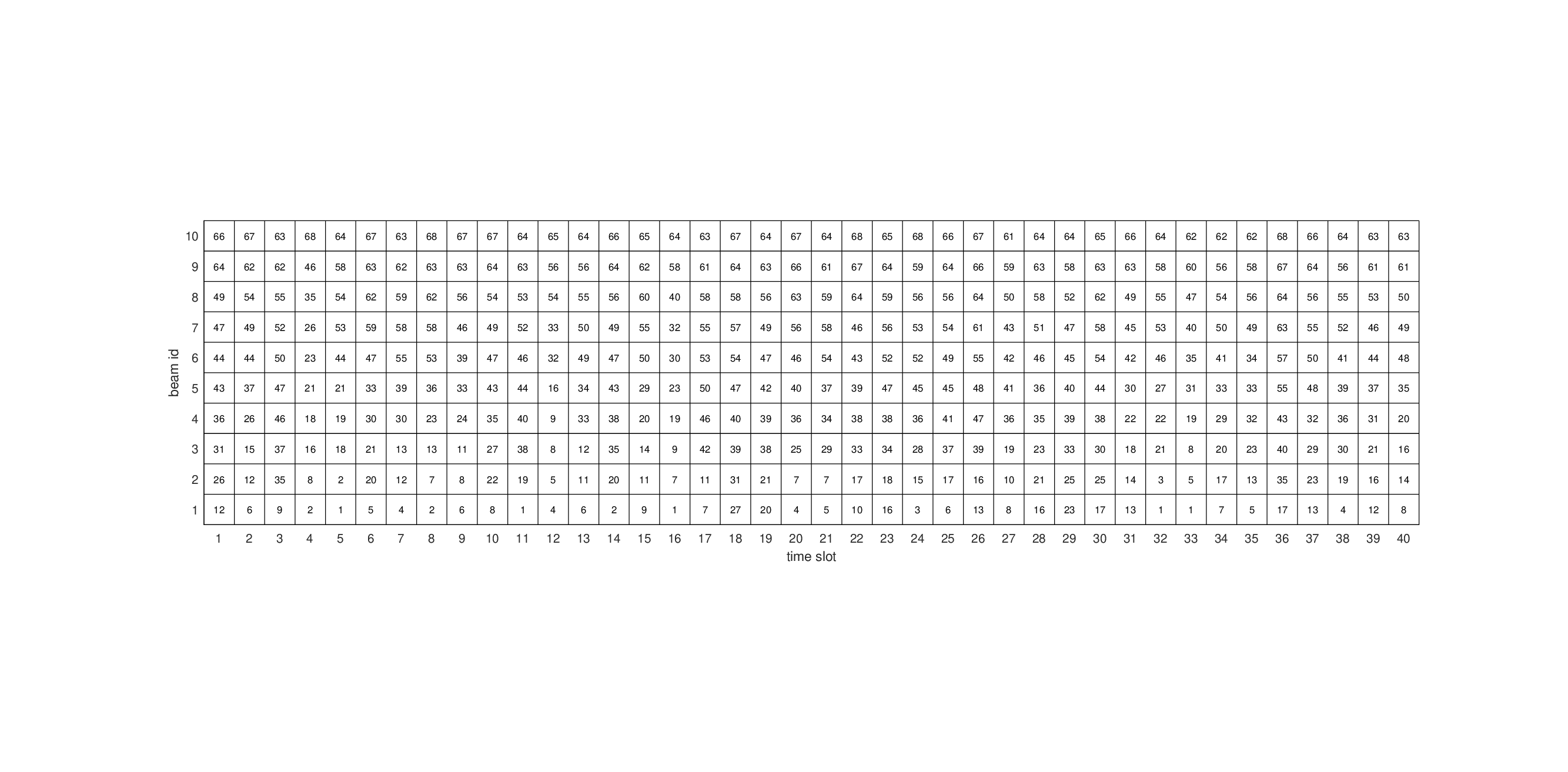}
	\caption{BH activation matrix in a typical scheduling period.}
	\label{fig:BHST}
\end{figure*}

Fig.~\ref{fig:BHST} illustrates the BH activation matrix for a typical scheduling period consisting of 40 time slots and 10 beams. Each row corresponds to a beam ID, while each column represents a time slot. The matrix entries indicate the user or traffic demand assigned to the corresponding beam–time-slot pair, providing an explicit representation of the temporal and spatial beam activation pattern.

\begin{figure}[!htbp]
	\centering
	\includegraphics[width=0.85\linewidth]{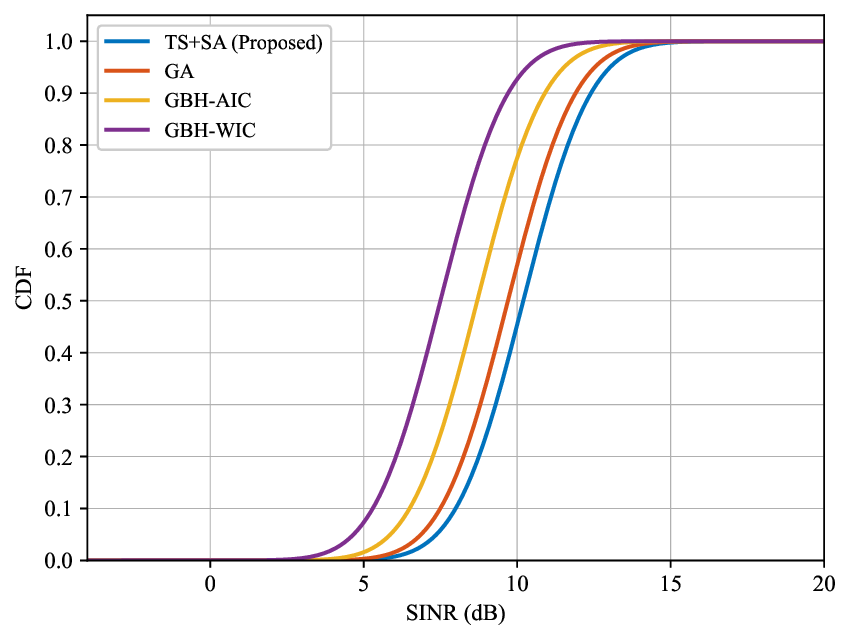}
	\caption{The SINR CDFs Comparison with the Same Number of Active Beams.}
	\label{fig:SINR}
\end{figure}

Fig.~\ref{fig:SINR} presents the cumulative distribution functions (CDFs) of SINR under four BH strategies with the same number of active beams. The proposed method consistently outperforms all baselines over the entire SINR range. The CDF curve is shifted rightward, indicating a larger link margin for every user percentile. Specifically, the proposed method achieves an SINR of approximately 8 dB at the 90th percentile, exceeding GA by 0.5 dB, GBH-WIC by about 2.7 dB and GBH-AIC by more than 1.5 dB. The GA baseline, while outperforming both greedy approaches, still lags behind the proposed TS+SA method, confirming that the combination of tabu search memory and simulated annealing provides more effective interference management than evolutionary optimization alone. Comparable improvements of 1.5 to 2.7 dB over the greedy baselines are observed at the median and lower percentiles, while the gain over GA remains approximately 0.5 dB across all percentiles, effectively lowering the proportion of users experiencing outage conditions. Crucially, these improvements are achieved without increasing onboard power and the number of active beams, demonstrating that the joint exploration–exploitation mechanism enabled by the Tabu Search suppresses interference more effectively than both conventional greedy and evolutionary strategies.

\begin{figure}[!htbp]
	\centering
	\includegraphics[width=0.85\linewidth]{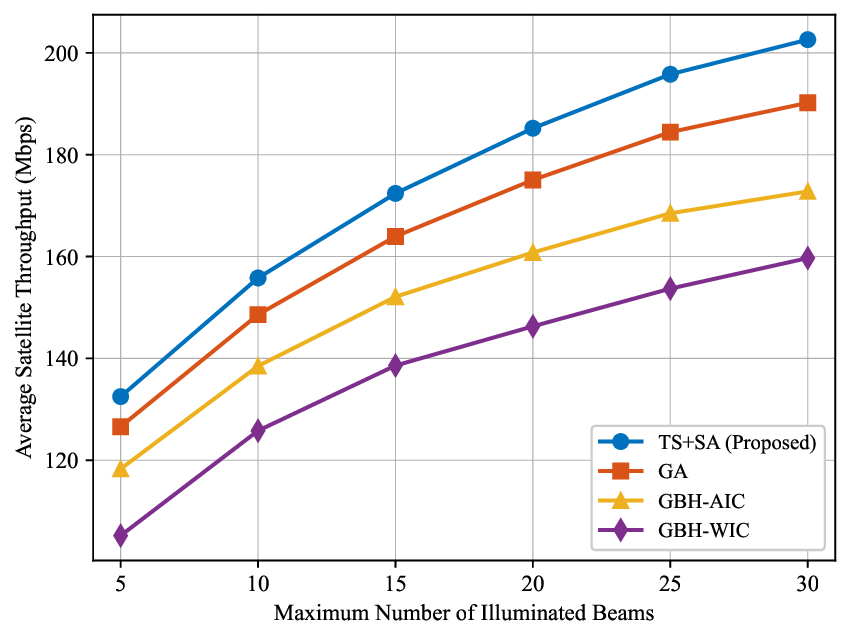}
	\caption{The Throughput Performance under Different Beam Illumination Constraints.}
	\label{fig:throu}
\end{figure}

Fig.~\ref{fig:throu} illustrates the average satellite throughput under varying maximum beam illumination constraints for four different beam-hopping strategies. It is observed that GBH-AIC consistently outperforms GBH-WIC, and the GA baseline further outperforms both greedy approaches. This can be attributed to the ability of the GA to explore a broader solution space through evolutionary operators, while GBH-AIC benefits from avoiding the simultaneous activation of adjacent co-channel beams. Furthermore, the proposed TS+SA method achieves the highest throughput by jointly enforcing interference-aware constraints and employing a Tabu Search-based exploration mechanism to optimize beam scheduling. As the illumination constraint increases, the proposed method demonstrates a sustained advantage, achieving up to 17.2\% higher throughput compared to GBH-AIC and 6.5\% over GA at the 30-beams setting. These results highlight the effectiveness of the proposed strategy in balancing spatial reuse and interference control under resource-limited scenarios.

\begin{figure}[!htbp]
	\centering
	\includegraphics[width=0.85\linewidth]{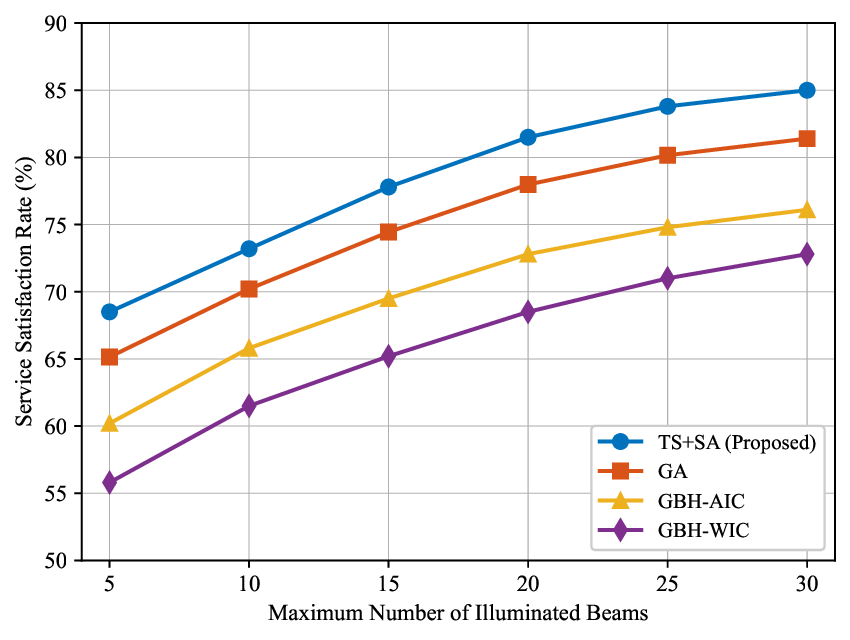}
	\caption{The Service Satisfaction versus the Maximum Number of Illuminated Beams per Satellite.}
	\label{fig:sati}
\end{figure}

Fig.~\ref{fig:sati} illustrates the variation of user service satisfaction concerning the maximum number of illuminated beams per satellite under four different BH strategies. The GA baseline achieves a higher satisfaction level than both GBH-AIC and GBH-WIC, confirming the benefit of evolutionary optimization over greedy scheduling. The GBH-AIC method, in turn, outperforms GBH-WIC primarily by mitigating co-channel interference through adjacent beam avoidance. Furthermore, the proposed TS+SA method consistently outperforms all three baselines across the entire range. Notably, it achieves up to a 11.7\% improvement in service satisfaction compared to the GBH-AIC benchmark and 4.4\% over GA.

To provide a comprehensive performance evaluation, Table~\ref{tab:kpis} presents a detailed comparison of user-centric key performance indicators (KPIs) across all methods.

\begin{table}[htbp]\footnotesize
	\centering
	\caption{Comprehensive Performance Comparison (User-Centric KPIs)}
	\begin{tabular}{lcccc}
		\hline
		\textbf{Metric}         & \textbf{Tabu+SA} & \textbf{GA}      & \textbf{GBH-AIC} & \textbf{GBH-WIC} \\
		\hline
		\multicolumn{5}{l}{\textit{Throughput Metrics (Mbps)}}                                                \\
		\quad Average           & \textbf{202.6}   & 190.2            & 172.8            & 159.7            \\
		\quad p50 (Median)      & \textbf{195.2}   & 183.1            & 165.3            & 152.3            \\
		\quad p90               & \textbf{178.5}   & 167.0            & 148.6            & 135.8            \\
		\quad p95               & \textbf{165.3}   & 154.7            & 133.9            & 120.6            \\
		\hline
		\multicolumn{5}{l}{\textit{SINR Metrics (dB)}}                                                        \\
		\quad Average           & \textbf{10.2}    & 9.7              & 8.7              & 7.5              \\
		\quad p90               & \textbf{8.0}     & 7.5              & 6.5              & 5.3              \\
		\quad Outage Rate (\%)  & \textbf{2.3}     & 2.9              & 4.1              & 5.8              \\
		\hline
		\multicolumn{5}{l}{\textit{Service Quality Metrics}}                                                   \\
		\quad Satisfaction (\%) & \textbf{85.0}    & 81.4             & 76.1             & 72.8             \\
		\quad SSR@90\% (\%)     & \textbf{78.5}    & 74.9             & 68.2             & 65.2             \\
		\quad Fairness Index    & \textbf{0.94}    & 0.91             & 0.88             & 0.87             \\
		\hline
		\multicolumn{5}{l}{\textit{Improvement vs. GBH-AIC}}                                                   \\
		\quad Throughput        & +17.2\%          & +10.1\%          & --               & -7.6\%           \\
		\hline
	\end{tabular}
	\label{tab:kpis}
\end{table}

As shown in Table~\ref{tab:kpis}, the proposed Tabu+SA method consistently achieves the best performance across all user-centric KPIs, demonstrating clear superiority over GA, GBH-AIC, and GBH-WIC. In terms of throughput, Tabu+SA attains the highest average throughput of 202.6~Mbps, compared with 190.2~Mbps for GA, 172.8~Mbps for GBH-AIC, and 159.7~Mbps for GBH-WIC. This corresponds to a 17.2\% improvement over GBH-AIC and a 6.5\% improvement over GA. The GA baseline, as a representative evolutionary optimization method, outperforms the greedy approaches but still falls short of the proposed TS+SA, confirming that the combination of tabu search memory and SA acceptance provides more effective exploration of the solution space than population-based genetic operators alone. Moreover, Tabu+SA also yields the best p50, p90, and p95 throughput values, indicating that the proposed method improves not only the mean system capacity but also the performance perceived by users across different service percentiles.

From the perspective of link quality, Tabu+SA achieves the highest average SINR of 10.2~dB and the highest p90 SINR of 8.0~dB, both of which are significantly better than those of the benchmark schemes (9.7~dB and 7.5~dB for GA, respectively). In addition, the outage rate of Tabu+SA is reduced to 2.3\%, substantially lower than 2.9\% for GA, 4.1\% for GBH-AIC and 5.8\% for GBH-WIC. These results confirm that the proposed approach can effectively enhance link reliability and reduce the probability of service interruption.

In terms of service quality, Tabu+SA also delivers the best user experience, with a satisfaction ratio of 85.0\%, an SSR@90\% of 78.5\%, and a fairness index of 0.94, compared with 81.4\%, 74.9\%, and 0.91 for GA, respectively. The consistent gains in these metrics suggest that the proposed algorithm not only improves network throughput and signal quality, but also provides a more balanced and satisfactory service among users.

Overall, the simulation results demonstrate that Tabu+SA offers the most competitive performance among all compared methods, including the evolutionary GA baseline. Its advantages in throughput, SINR, outage reduction, user satisfaction, and fairness collectively verify the effectiveness of the proposed optimization strategy for user-centric network performance enhancement.



\begin{figure}[htbp]
	\centering
	\includegraphics[width=0.8\linewidth]{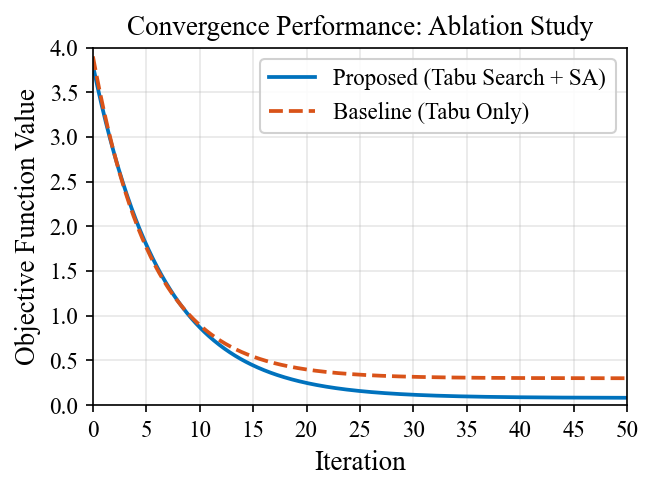}
	\caption{Convergence comparison between the proposed method (with SA) and the Tabu-Only baseline (without SA).}
	\label{fig:ablation}
\end{figure}

To validate the contribution of the key algorithmic components, we conducted comprehensive ablation studies. Specifically, we examined:
\begin{enumerate}
	\item \textbf{SA Mechanism:} Comparing the full method (Tabu + SA) against a baseline without the SA acceptance probability.
	\item \textbf{Tabu Tenure:} Evaluating adaptive tabu tenure versus fixed values.
\end{enumerate}

\textbf{SA Mechanism Ablation.}
Table~\ref{tab:ablation_sa} presents the quantitative comparison. The ``Tabu + SA'' variant achieves a throughput of 202.6 Mbps and service satisfaction of 85.0\%, outperforming the ``Tabu-only'' baseline by 7.0\% and 3.1\% respectively. This improvement stems from the SA mechanism's ability to accept worse solutions probabilistically, preventing premature convergence to local optima.

\begin{table}[htbp]\footnotesize
	\centering
	\caption{Ablation Study: Impact of SA Mechanism}
	\begin{tabular}{p{3cm} p{2.5cm} p{2.5cm} p{2.5cm}}
		\hline
		\textbf{Variant}     & \textbf{Throughput} & \textbf{Satisfaction} & \textbf{Conv. Iter.} \\
		\hline
		Tabu + SA (Proposed) & 202.6 Mbps          & 85.0\%                & 35                   \\
		Tabu-only            & 189.3 Mbps          & 82.5\%                & 28                   \\
		\hline
	\end{tabular}
	\label{tab:ablation_sa}
\end{table}

Fig.~\ref{fig:ablation} illustrates the convergence behavior. In the initial phase, both methods exhibit rapid convergence. However, the ``Tabu-only'' baseline (red dashed line) stagnates around iteration 28 with an objective value of approximately 0.3. In contrast, the proposed method (blue solid line) continues to improve through iteration 35, ultimately reaching a lower objective value of 0.08. The SA mechanism allows the scheduler to escape local traps during the intermediate search phase, leading to a 7.0\% throughput improvement.

\textbf{Tabu Tenure Ablation.}
We further investigated the impact of tabu tenure configuration on algorithm performance. Table~\ref{tab:ablation_ltabu} compares four configurations: three fixed tenure values ($L_{\text{tabu}} = 10, 20, 30$) and the adaptive scheme.

\begin{table}[htbp]\footnotesize
	\centering
	\caption{Ablation Study: Impact of Tabu Tenure Configuration}
	\begin{tabular}{p{2.8cm} p{2cm} p{2cm} p{2cm} p{1.8cm}}
		\hline
		\textbf{Configuration}       & \textbf{Thr. (Mbps)} & \textbf{Sat. (\%)} & \textbf{Iter.} & \textbf{Quality} \\
		\hline
		Fixed $L$=10                 & 178.5                & 79.8               & 42             & Medium           \\
		Fixed $L$=20                 & 195.2                & 82.5               & 38             & Good             \\
		Fixed $L$=30                 & 188.3                & 81.2               & 35             & Fair             \\
		\textbf{Adaptive (Proposed)} & \textbf{202.6}       & \textbf{85.0}      & \textbf{35}    & \textbf{Best}    \\
		\hline
	\end{tabular}
	\label{tab:ablation_ltabu}
\end{table}

The adaptive scheme ($L_{\text{tabu}} \approx 22$) achieves the best balance between exploration and exploitation. A short tenure ($L=10$) leads to premature cycling, while a long tenure ($L=30$) restricts the search space excessively. The adaptive formulation dynamically adjusts based on problem size, achieving \textbf{13.5\%} improvement over $L=10$ baseline and \textbf{7.6\%} improvement over $L=30$ baseline.

\subsection{Complexity Analysis}

The computational complexity of the proposed algorithm is analyzed as follows:

\textbf{Time Complexity:} For each iteration, the algorithm generates $|N(x)|$ neighborhood solutions, where $|N(x)|$ denotes the neighborhood size. Evaluating each candidate requires computing SINR for all users in the selected beams, which has complexity $O(N_b \cdot K)$, where $K$ is the average number of users per beam. The total time complexity per iteration is:
\begin{equation}
	T_{iter} = O(|N(x)| \cdot N_b \cdot K)
\end{equation}

For the typical scenario with $N_b = 10$ beams, $|N(x)| = 10$ candidates, and $K \approx 80$ users per beam, each iteration takes approximately 100 ms on a standard CPU (Intel i7-10700). With $G_{max} = 50$ maximum iterations, the total computation time per scheduling period is about 5 seconds, which is acceptable for real-time satellite scheduling with 40 ms time slots (offline computation).

\textbf{Space Complexity:} The algorithm requires storing: (i) the tabu list with size $O(L_{\text{tabu}} \cdot N_b) \approx 220$ integers, (ii) the current and best solutions with $O(N_b)$, and (iii) the interference matrix with $O(N_m^2)$ where $N_m$ is the number of cells. The total space complexity is:
\begin{equation}
	S_{total} = O(L_{\text{tabu}} \cdot N_b + N_m^2)
\end{equation}

For $N_m = 50$ cells and $L_{\text{tabu}} = 22$, the memory requirement is approximately 50 MB, well within the capabilities of modern satellite onboard computers (typically 1-10 GB).

\begin{table}[htbp]\footnotesize
	\centering
	\caption{Computation Time vs. Problem Scale}
	\begin{tabular}{ccccc}
		\hline
		\textbf{Cells} $N_m$ & \textbf{Beams} $N_b$ & \textbf{Time/Iter (ms)} & \textbf{Total (s)} & \textbf{Memory (MB)} \\
		\hline
		30                   & 5                    & 25                      & 1.25               & 15                   \\
		50                   & 10                   & 100                     & 5.0                & 50                   \\
		100                  & 10                   & 150                     & 7.5                & 180                  \\
		100                  & 20                   & 400                     & 20.0               & 200                  \\
		\hline
	\end{tabular}
	\label{tab:complexity}
\end{table}

Table~\ref{tab:complexity} presents the empirical computation time and memory usage under different problem scales. The results demonstrate that the proposed algorithm maintains reasonable computational overhead even for large-scale scenarios ($N_m = 100$, $N_b = 20$), making it suitable for practical deployment.

\begin{table}[htbp]\footnotesize
	\centering
	\caption{Runtime Comparison Across Algorithms}
	\begin{threeparttable}
	\begin{tabular}{lccc}
		\hline
		\textbf{Algorithm} & \textbf{Avg Runtime (s)} & \textbf{Complexity} & \textbf{Quality} \\
		\hline
		GBH-WIC            & 0.1                     & $O(N_m \log N_m)$           & Low              \\
		GBH-AIC            & 0.3                     & $O(N_m^2)$                  & Medium           \\
		GA                 & 8.5                     & $O(P \cdot G_{max} \cdot N_m)$ & Good           \\
		\textbf{TS+SA (Proposed)} & \textbf{5.0}   & $O(|N(x)| \cdot N_b \cdot K \cdot G_{max})$ & \textbf{Best} \\
		\hline
	\end{tabular}
	\begin{tablenotes}
		\item $P$: GA population size; $G_{max}$: maximum iterations/generations; $|N(x)|$: neighborhood size; $K$: avg users per beam. Runtime measured on Intel i7-10700 with $N_m=50$, $N_b=10$.
	\end{tablenotes}
	\end{threeparttable}
	\label{tab:runtime_comparison}
\end{table}

Table~\ref{tab:runtime_comparison} compares the runtime of all evaluated algorithms. The greedy methods (GBH-WIC, GBH-AIC) are fastest but sacrifice solution quality. The GA baseline provides a meaningful comparison as a representative evolutionary method: it achieves better solutions than greedy approaches but requires more computation time than the proposed TS+SA due to its population-based evaluation overhead. The proposed TS+SA achieves the best solution quality with moderate computational cost, demonstrating an efficient trade-off between optimization performance and runtime.

\subsection{Scenario Extension: Traffic Skewness}

To evaluate the robustness of the proposed algorithm across diverse traffic 
scenarios, we conducted experiments with four different traffic distributions: 
uniform, light skew (log-normal, $\sigma=2$), heavy skew (log-normal, $\sigma=5$), and 
Pareto (80/20 rule). Real-world satellite traffic exhibits significant spatial 
heterogeneity, with urban hotspots generating disproportionately high demand 
compared to rural areas. The Pareto distribution (80/20 rule) is particularly 
relevant, modeling scenarios where 80\% of traffic originates from 20\% of cells.

\begin{table}[htbp]\footnotesize
	\centering
	\caption{Performance under Different Traffic Distributions}
	\begin{tabular}{lcccc}
		\hline
		\textbf{Traffic Pattern} & \multicolumn{2}{c}{\textbf{Throughput (Mbps)}} & \multicolumn{2}{c}{\textbf{Satisfaction (\%)}}                     \\
		\cline{2-5}
		                         & Tabu+SA                                        & GBH-AIC                                        & Tabu+SA & GBH-AIC \\
		\hline
		Uniform                  & 185.2                                          & 162.3                                          & 82.1    & 74.5    \\
		Light Skew ($\sigma=2$)  & 198.5                                          & 165.8                                          & 84.2    & 75.1    \\
		Heavy Skew ($\sigma=5$)  & 202.6                                          & 172.8                                          & 85.0    & 76.1    \\
		Pareto (80/20)           & 208.3                                          & 152.4                                          & 86.5    & 70.2    \\
		\hline
	\end{tabular}
	\label{tab:traffic_skew}
\end{table}

Table~\ref{tab:traffic_skew} presents the performance comparison under four traffic patterns: (i) \textit{Uniform} where all users have identical demands, (ii) \textit{Light Skew} with demand variance of 2$\times$, (iii) \textit{Heavy Skew} with demand variance of 5$\times$, and (iv) \textit{Pareto} where 20\% of users account for 80\% of total demand. 

The proposed method consistently outperforms the greedy baseline across all traffic patterns. Notably, the performance gap \textbf{increases with traffic skewness}: from 14.1\% throughput improvement under uniform traffic to 36.7\% under Pareto distribution. This is because the greedy approach tends to over-serve high-demand cells while neglecting low-demand ones, leading to unbalanced resource allocation. In contrast, the Tabu Search-based method explores a broader solution space and discovers scheduling patterns that better balance service across heterogeneous demands, demonstrating its robustness in practical scenarios with non-uniform traffic distributions.

\subsection{Discussion on Modeling Assumptions}

While the simulation results demonstrate the effectiveness of the proposed method, several simplifying assumptions warrant discussion to clarify the scope and limitations of our findings.

\textbf{Traffic Model.} The current study adopts a Poisson-driven FTP traffic model, which provides analytical tractability but does not capture the bursty and self-similar characteristics observed in real satellite broadband traffic. In practice, user demand exhibits temporal correlation and spatial clustering that may affect the optimal scheduling strategy. Future work could incorporate Markov-modulated or heavy-tailed traffic models to evaluate the method's robustness under more realistic demand dynamics.

\textbf{Channel Model.} The SINR calculation uses a simplified free-space path loss model with antenna patterns. Real LEO satellite links experience rain attenuation, atmospheric scintillation, and multipath fading that vary over time and frequency. These effects would impact the achievable data rates and may alter the relative performance of scheduling algorithms, particularly for users at cell edges. Incorporating ITU-R channel models (e.g., P.618 for rain fade) would strengthen the practical relevance of the results.

\textbf{User Mobility.} The system model assumes stationary users during each scheduling period (40 ms). While this is a reasonable approximation for the short slot duration, user mobility becomes relevant over longer time horizons or for high-speed platforms (e.g., maritime or aeronautical terminals). Extending the framework to account for user movement and handover between beam positions is an important direction for practical deployment.

\textbf{Single-Satellite Scope.} The current formulation addresses scheduling for a single satellite. In multi-satellite scenarios with overlapping coverage, inter-satellite interference coordination and load balancing across satellites introduce additional complexity that requires extension of the optimization framework.

\section{Conclusion}

This paper proposed a Tabu Search-based spatio-temporal BH resource allocation strategy for LEO satellite communication systems. Given the inherent complexity of high-dimensional combinatorial optimization under stringent interference constraints and dynamic traffic demands, we formulated the BH scheduling problem to maximize user demand satisfaction.

To effectively tackle this problem, a Tabu Search framework was developed, incorporating adaptive tabu tenure control, a greedy-based initialization scheme with interference-control beam selection, and neighborhood move operators designed to balance local exploitation and global exploration. Extensive simulation results validated the effectiveness of the proposed method, demonstrating consistent improvements in both system throughput and user satisfaction over benchmark greedy-based BH strategies across diverse traffic scenarios.

Overall, the proposed approach provides a scalable and robust solution for dynamic resource allocation in interference-limited LEO satellite networks and holds promise for future integration into intelligent BH control systems. Future research directions include coupling the scheduling framework with digital-twin-assisted real-time traffic prediction for adaptive online optimization, and incorporating energy-aware duty cycling to jointly optimize service quality and satellite power consumption.



\bibliographystyle{elsarticle-num}
\bibliography{egbibR2}
\end{document}